\documentclass[aps,prb,amsmath,twocolumn]{revtex4}

\usepackage{graphicx}
\usepackage{bm}

\begin{document}

\title{Fluctuations of the Josephson current and electron-electron
interactions in superconducting weak links}

\author{Artem V. Galaktionov$^1$ and Andrei D.
Zaikin$^{2,1}$} \affiliation{$^1$I.E. Tamm Department of
Theoretical Physics, P.N. Lebedev Physics Institute, 119991
Moscow, Russia} \affiliation{$^2$ Institute for Nanotechnology,
Karlsruhe Institute of Technology (KIT), 76021 Karlsruhe, Germany}

\begin{abstract}
We derive a microscopic effective action for superconducting contacts with arbitrary transmission distribution of conducting channels. Provided fluctuations of the Josephson phase remain sufficiently small our formalism allows to fully describe fluctuation and interaction effects in such systems. As compared to the well studied tunneling limit our
analysis yields a number of qualitatively new features which occur
due to the presence of subgap Andreev bound states in the system. We investigate
the equilibrium  supercurrent noise and evaluate the electron-electron
interaction correction to the Josephson current across superconducting
contacts. At $T=0$ this correction is found to vanish for fully transparent contacts indicating the absence of Coulomb effects in this limit.

\end{abstract}


\maketitle

\section{Introduction}
It is well known that supercurrent can flow through a non-superconducting
barrier between two superconducting reservoirs. Initially this effect was
predicted \cite{BJ} and microscopically analyzed \cite{AB} for a specific
case of (usually very thin) tunnel insulating barriers. Later it was understood that non-dissipative transport of Cooper pairs between two superconductors is also possible in many other types of weak links, such as, e.g., quantum point contacts \cite{KO} and superconductor-normal-metal-superconductor ($SNS$) junctions \cite{SNScl,SNSd}, i.e. if a piece of a normal metal is placed in-between two superconductors. In contrast to tunnel junctions, in $SNS$ systems at sufficiently low temperatures appreciable supercurrent can flow even though  a normal layer can be as thick as few microns.

It turned out that the Josephson effect in superconducting weak links without tunnel barriers is directly related to another fundamentally important phenomenon: Andreev reflection \cite{AR}. Suffering Andreev reflections at both $NS$ interfaces, quasiparticles with energies below the superconducting gap are effectively ``trapped'' inside the junction forming a discrete set of levels which can be tuned by passing the supercurrent across the system. At the same time, these subgap Andreev levels themselves contribute to the supercurrent  thus making the behavior of superconducting point contacts and $SNS$ junctions in many respects different from that of tunnel barriers. For an extended review summarizing various features of dc Josephson effect in different types of superconducting weak links we refer the reader to Refs. \onlinecite{lam,bel,SaMiZhe}.

 The number of Cooper pairs transferred between two superconductors and, hence, the Josephson current can fluctuate around its mean value \cite{Madrid,Averin,Yip}.  While at non-zero temperatures thermal fluctuations of the supercurrent should naturally exist in all types of weak links, in the limit $T \to 0$ the relevant physics is essentially determined by the presence or absence of subgap Andreev states. Provided
 such states are present fluctuations of the supercurrent do in general occur even at $T =0$ and at subgap frequencies. E.g., the equilibrium supercurrent correlation functions show pronounced peaks at frequencies equal to the distance between Andreev levels inside the weak link. The amplitudes of such peaks turn out to scale as \cite{Madrid} $\sum_nT_n^2(1-T_n)$, where $T_n$ is the normal transmission of the $n$-th conducting mode of the barrier and the sum is taken over all such modes. The latter dependence implies that ground state fluctuations of the supercurrent can be expected neither in the limit of low barrier transmissions $T_n \to 0$ (i.e. in tunnel barriers where no Andreev states are present) nor in fully open contacts with  $T_n \to 1$.

Note that the above considerations remain applicable if one can neglect Coulomb effects. In small-size superconducting contacts, however, such effects can be
important and should in general be taken into account. A lot is known about interplay between fluctuations
and charging effects in superconducting tunnel barriers \cite{SZ}. Here we examine the properties of superconducting junctions going beyond the tunneling limit. We will analyze fluctuation and interaction effects and demonstrate that Coulomb blockade in such junctions weakens with increasing barrier transmissions and eventually disappears in the limit of fully open superconducting contacts.

The structure of our paper is as follows. In Sec. II we derive an effective action for superconducting
contacts with arbitrary distribution of channel transmissions which enables one to describe equilibrium fluctuations of the current and interaction effects.   In Sec. III we make use of this action and evaluate the supercurrent noise in superconducting contacts. Low frequency current response and capacitance renormalization due to retardation effects are discussed in Sec. IV. In Sec. V we analyze the interaction correction to the Josephson current. A brief summary of our main results is presented in Sec. VI. Some general expressions and technical details are relegated to Appendix.

\section{Effective action and phase fluctuations}

In what follows we will adopt the standard model of a
superconducting contact and consider two big superconductors
connected with each other via a normal conductor (see Fig. 1)
characterized by arbitrary transmission distribution $T_n$ of its
spin-degenerate conducting channels. Below we will only consider
the limit of sufficiently short normal conductors with effective
Thouless energy $\varepsilon_{\rm Th}$ strongly exceeding the
superconducting gap $\Delta$ in both reservoirs, $\varepsilon_{\rm
Th} \gg \Delta$. In addition, the normal conductor length is
assumed to be much shorter than dephasing and inelastic relaxation
lengths. Coulomb interaction between electrons in the contact area
is described in a standard manner by an effective capacitance $C$.

\begin{figure}
\includegraphics[width=6cm]{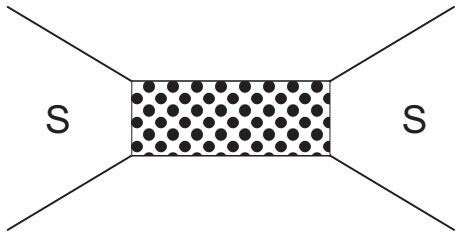}
\caption{Short coherent conductor between two superconducting reservoirs.}
\end{figure}

We will assume that the contact is biased by external current $I$
which does not exceed the critical one $I_C$ and, hence, can flow
through the contact without any dissipation, i.e. $I \equiv I_S$. In the absence of fluctuations this external current sets the value of the order parameter phase difference $\chi =\chi_L-\chi_R$ between two superconductors. The corresponding implicit dependence of $\chi$ on the supercurrent $I_S$ has the form \cite{KO}
\begin{eqnarray}
&& I_S(\chi )=\frac{e\Delta\sin\chi}{2}\sum_n\frac{T_n}{\sqrt{1-T_n\sin^2(\chi/2)}}
\label{Ichi}
\\ && \times \tanh\frac{\Delta\sqrt{1-T_n\sin^2(\chi/2)}}{2T}.
\nonumber
\end{eqnarray}
Here and below we set $\hbar =1$ and define the electron charge to be $-e$.

In order to analyze fluctuation and interaction effects in such
superconducting contacts we will allow for fluctuations of the superconducting phase difference around its average value
$\chi$ and employ the effective action
formalism combined with the scattering matrix technique. This
approach was proven to be very successful in the case of normal
conductors \cite{GZ01,GGZ03,KN,GZ04} and NS hybrid structures
\cite{GZ06,GZ09}. Following the usual procedure we express the kernel
$J$ of the evolution operator on the Keldysh contour in terms of a
path integral over the fermionic fields which can be integrated
out after the standard Hubbard-Stratonovich decoupling of the
interacting term \cite{SZ}. Then the kernel $J$ acquires the form
\begin{equation}
J=\int {\cal D} \varphi_1{\cal D}\varphi_2\exp(iS_c[\varphi
]+iS_t[\varphi ]), \label{pathint}
\end{equation}
where the terms $S_c[\varphi ]$ and $S_t[\varphi ]$ account
respectively for charging effects and for the transfer of
electrons and Cooper pairs between two superconducting reservoirs.
Both these terms represent the functionals of the fluctuating
phase variables $\varphi_{1,2}(t)$ defined on the forward and backward parts of the Keldysh contour and related to fluctuating voltages
$V_{1,2}$ across the conductor as $\dot\varphi_{1,2}(t)=eV_{1,2}$.
With the aid of the Josephson relation one trivially identifies the superconducting phase difference on two branches
of the Keldysh contour as $\chi +2\varphi_{1,2}(t)$.

The charging term is taken in the standard form \cite{SZ}
\begin{eqnarray}
S_c[\varphi]= \frac{C}{2e^2}\int\limits_0^t dt'
(\dot\varphi_{1}^2-\dot\varphi_{2}^2)\equiv
\frac{C}{e^2}\int\limits_0^t dt
\dot\varphi^+\dot\varphi^-,
\label{Sc}
\end{eqnarray}
where we also introduced ``classical'' and ``quantum'' parts of
the phase, respectively $\varphi_+=(\varphi_1+\varphi_2)/2$ and
$\varphi_-=\varphi_1-\varphi_2$. The structure of the term
$S_t[\varphi ]$ is the same as in the normal case, one should only
replace normal propagators by $2\times2$ Green-Gorkov matrix
functions \cite{SZ,Z,SN}. The corresponding result can be
expressed in the form \cite{SN}
\begin{equation}
S_t[\varphi]=-\frac{i}{2}\sum_n{\rm Tr} \ln \left[
1+\frac{T_n}{4}\left( \left\{ \check Q_L(\varphi ), \check Q_R
\right\}-2\right) \right], \label{St}
\end{equation}
where $\check Q_{L,R}$ are $4\times4$ Green-Keldysh matrices of
the left and right superconducting electrodes. The product of
these matrices implies time convolution and curly brackets denote
anticommutation.

Without loss of generality we can set the electric potential (and,
hence, fluctuating phases) of the right superconducting terminal
equal to zero. Then the Green-Keldysh matrix of this electrode can
be written in a simple form
\begin{equation}
\check Q_R=\check g=\left( \begin{array}{cc} \hat g^R& \hat g^K\\0
& \hat g^A\end{array}\right),
\end{equation}
where $\hat g^{R,A}$ are retarded and advanced $2\times2$ matrix
functions
\begin{equation}
\hat g^{R,A}=\frac{\epsilon \hat\tau_3+\hat \Delta}{\xi^{R,A}},
\quad \xi^{R,A}=\pm\sqrt{(\epsilon\pm i\delta)^2-\Delta^2}
\end{equation}
and $\hat g^K=\hat g^R F-F \hat g^A$ is the Keldysh matrix, where
$F(\epsilon)=\tanh (\epsilon /2T)$ is the Fourier transform of
$F(t)=-iT/\sinh [\pi T t]$ and $\hat\tau_i$ are the Pauli
matrices. For simplicity we choose the order parameter $\Delta$ of
the right superconductor real and, hence, we can set $\hat \Delta
=i\Delta \hat\tau_2$. In order to properly account for analytic
properties of the functions $\xi^{R,A}$ here we keep an
infinitesimally small imaginary part $i\delta$ which allows to
define $\xi^{R,A}=\pm {\rm
sgn}\,\epsilon\sqrt{\epsilon^2-\Delta^2}$ for $|\epsilon|>\Delta$
and $\xi^{R,A}=i\sqrt{\Delta^2-\epsilon^2}$ for
$|\epsilon|<\Delta$.

The Green-Keldysh matrix $\check Q_L$ of the left superconducting
electrode reads
\begin{equation}
\check Q_L(\varphi)(t,t')=\check {\cal L} \check {\cal
M}_+(t)\check {\cal L} \check g(t,t') \check {\cal L} \check {\cal
M}_-(t')\check {\cal L},
\end{equation}
where we defined the matrices
\begin{equation}
\check {\cal L}=\frac{1}{\sqrt{2}}\left( \begin{array}{cc}\hat 1&
\hat 1\\ \hat 1&-\hat 1 \end{array}\right) \nonumber
\end{equation}
and
\begin{eqnarray}
&& \check{\cal M}_{\pm}= \nonumber\\
&&\left(
\begin{array}{cc} \exp\left[\pm i\left(\frac{\chi}{2}+\varphi_1(t)\right)\hat\tau_3\right] & 0 \\  0 &
 \exp\left[\pm i\left(\frac{\chi}{2}+\varphi_2(t)\right)\hat\tau_3\right]\end{array}\right).
\nonumber
\end{eqnarray}
Substituting the above expressions for $\check Q_L$ and $\check
Q_R$ into Eq. (\ref{St}) we arrive at the action which fully
describes transfer of electrons and Cooper pairs to all orders in
$T_n$. In the case of tunnel barriers the channel transmissions
remain small and one can expand $S_t$ in powers of $T_n$. Keeping
the lowest order terms $\sim T_n$ of this expansion one recovers
the well-known Ambegaokar-Eckern-Sch\"on (AES) action \cite{SZ}.
Here, however, we will go beyond the tunneling limit and analyze
fluctuation effects at arbitrary transmission values $T_n$.

To this end we will proceed similarly to Ref. \onlinecite{GZ09}
and introduce the matrix
\begin{equation}
\check X_0=1-T_n/2+(T_n/4)\left\{ \check Q_L, \check
Q_R \right\}|_{\varphi_{\pm}=0}.
\end{equation}
As the action $S_t$ vanishes for $\varphi_-(t)=0$ one has ${\rm
Tr}\ln \check X_0=0$. Making use of this property we can
identically transform the action (\ref{St}) to
\begin{equation}
S_t=-\frac{i}{2}\sum_n{\rm Tr} \ln \left[ 1+\check X_0^{-1}\circ
\check X' \right], \label{xx}
\end{equation}
where
\begin{equation}
\check X'=1+(T_n/4)\left( \left\{ \check Q_L, \check Q_R
\right\}-2\right)-\check X_0.
\end{equation}
With the aid of the
above expressions for $\check Q_{L,R}$ we obtain
\begin{equation}
\check X_0^{-1}=\left(\begin{array}{cc} \frac{ \xi^2\hat
1}{(\epsilon+i\delta)^2-\epsilon_n^2}&\frac{-i\pi\xi^2\hat
1}{\epsilon_n}\tanh\frac{\epsilon_n}{2T} \sum_{\pm}
\delta(\epsilon\pm\epsilon_n)
\\ 0&\frac{\xi^2\hat
1}{(\epsilon-i\delta)^2-\epsilon_n^2}
\end{array}\right), \label{x0}
\end{equation}
where $\xi^2=\epsilon^2-\Delta^2$ and the subgap Andreev level inside the contact
with energies $\pm\epsilon_n(\chi)$ are defined in a usual way as
\begin{equation}
\epsilon_n(\chi)=\Delta\sqrt{1-T_n\sin^2(\chi/2)}.
\label{And}
\end{equation}

Now let us assume that fluctuating phases $\varphi_\pm (t)$
(or fluctuating voltages) at the junction are sufficiently small
and perform regular expansion of the exact effective action in powers
of these phases. Expanding  $\check Q_L (\varphi)$ up to the second order in $\varphi_\pm$
(thus finding the matrix $\check X'$), from Eq. (\ref{xx}) we obtain
\begin{equation}
iS_t=-\frac{i}{e}\int\limits_0^t dt'I_S(\chi)\varphi_-(t')+ iS_R-S_I, \label{finalsS}
\end{equation}
where the supercurrent $I_S(\chi)$ is defined in Eq. (\ref{Ichi}) and
\begin{eqnarray}
S_R&=&\int\limits_0^t dt'\int\limits_{0}^{t}dt''
{\cal R}(t'-t'') \varphi^-(t')\varphi^+(t''), \label{SRRR}\\
S_I&=&\int\limits_{0}^{t}dt'\int\limits_{0}^{t}dt''
{\cal I}(t'-t'') \varphi^-(t')\varphi^-(t'') \label{SIII}
\end{eqnarray}
with both kernels ${\cal R}(t)$ and ${\cal I}(t)$ being real functions. The general expressions for these functions turn out to be somewhat lengthy and for this reason are presented in Appendix. Here we only emphasize some of the properties of ${\cal R}(t)$ and ${\cal I}(t)$.

To begin with, it is straightforward to verify that in the lowest order in barrier
transmissions $T_n$ the result (\ref{finalsS})-(\ref{SIII}) reduces to the standard AES action \cite{SZ} for tunnel barriers
in the limit of small phase fluctuations. Qualitatively new features emerge in higher orders in $T_n$ being directly related to the presence of subgap Andreev levels $\pm\epsilon_n(\chi)$ inside the contact. Consider, for instance, the kernel ${\cal I}(t)$ defined in Eq. (\ref{iw}). It can be split into three contributions of  different physical origin
\begin{equation}
 {\cal I}(t)={\cal I}_1(t)+{\cal I}_2(t)+{\cal I}_3(t).
 \end{equation}
The first of these terms, ${\cal I}_1(t)$, represents the subgap contribution due to discrete Andreev states. The Fourier transform of this term has the form (cf. the first line in Eq. (\ref{iw}))
\begin{eqnarray}
&&{\cal I}_{1\omega}=\frac{\pi\Delta^4}{4}\sum_n \bigg\{  \frac{ T_n^2\sin^2\chi}{2\epsilon_n^2(\chi)\cosh^2(\epsilon_n(\chi)/2T)}\delta(\omega)
\nonumber\\
&&+\frac{ T_n^2(1-T_n)\sin^4(\chi/2)}{\epsilon^2_n(\chi)}\left[1+\tanh^2(\epsilon_n(\chi)/2T)\right]\nonumber\\
&&\times [\delta\left(\omega-2\epsilon_n(\chi)\right)+ \delta\left(\omega+2\epsilon_n(\chi)\right)] \bigg\}.
\label{I1}
 \end{eqnarray}
It is obvious that this contribution is not contained in the AES action at all. The general expression for the second term  ${\cal I}_2(t)$ is defined by the second and third lines of Eq. (\ref{iw}). In the limit
of small barrier transmissions this term scales as ${\cal I}_2\propto T_n^{3/2}$ and, hence, is not contained in the AES action either. This
contribution can be interpreted as the ''interference term'' between
subgap Andreev levels and quasiparticle states above the gap. In the low
temperature limit $T \to 0$ the Fourier transform of this term ${\cal I}_{2\omega}$ differs from zero only at sufficiently high frequencies $|\omega |>\Delta +\epsilon_n(\chi)$. At higher temperatures $T \gtrsim \epsilon_n(\chi)$, however, ${\cal I}_{2\omega}$ vanishes only for $|\omega |<\Delta -\epsilon_n(\chi )$ and remains non-zero otherwise.
Finally, the third term ${\cal I}_3(t)$ accounts for the contribution of quasiparticles with energies above the gap. The Fourier transform of this term   ${\cal I}_{3\omega}$ is defined by the fourth and fifth lines of Eq. (\ref{iw}). In the high frequency limit $\omega \gg \Delta$ or for
$\Delta \to 0$ this term reduces to the standard result for a normal conductor
\begin{equation}
{\cal I}_{3\omega}\to\frac{\omega}{2e^2R_N}\coth\frac{\omega}{2T},
\label{R2}
\end{equation}
where $R_N$ is the normal contact resistance determined by the Landauer formula
\begin{equation}
\frac{1}{R_N}=\frac{e^2}{\pi}\sum_n T_n.
\end{equation}

Turning now to the function ${\cal R}(t)$ in  Eq. (\ref{SRRR}) we note that its Fourier transform can be represented as
${\cal R}_\omega={\cal R}'_\omega+i {\cal R}''_\omega$, where both ${\cal R}'_\omega$ and  ${\cal R}''_\omega$ are real functions. The function ${\cal R}'_\omega$ is even in $\omega$ while ${\cal R}''_\omega$ is an odd function of $\omega$, thus implying that the function ${\cal R}(t)$ is real.

The functions ${\cal R}(t)$ and ${\cal I}(t)$ are not independent. For instance, the Fourier transform ${\cal R}''_\omega$ is related to ${\cal I}_\omega$ by means of the fluctuation-dissipation relation
\begin{equation}
{\cal R}''_\omega=2 {\cal I}_\omega \tanh\frac{\omega}{2T}.
\label{FDT}
\end{equation}
The two functions ${\cal R}'_\omega$ and ${\cal R}''_\omega$ are in turn linked to each other by the causality principle: the function ${\cal R}(t)$ should vanish for $t<0$. The general expression for ${\cal R}'_\omega$ (\ref{im1}) and further details are presented in Appendix.

Finally we would like to point out that with the aid of the above Gaussian effective action one can
easily evaluate the phase-phase correlation functions for our problem.
Combining Eqs. (\ref{finalsS})-(\ref{SIII}) with (\ref{Sc}) one finds
\begin{eqnarray}
&& \langle \varphi_+(t_1)\varphi_+(t_2)\rangle= \label{phiphi}\\ &&
-\int\limits_{-\infty}^{\infty} \frac{d\omega}{2\pi} {\rm Im}\left(\frac{1}{C\omega^2/e^2+{\cal R}_\omega} \right)\coth \frac{\omega}{2 T}e^{-i\omega(t_1-t_2)},\nonumber\\ && \langle \varphi_+(t_1)\varphi_-(t_2)\rangle=\nonumber\\ &&
i\int\limits_{-\infty}^{\infty} \frac{d\omega}{2\pi} \left(\frac{1}{{C\omega^2/e^2+\cal R}_\omega} \right) e^{-i\omega(t_1-t_2)},\nonumber
\\ && \langle \varphi_-(t_1)\varphi_-(t_2)\rangle=0. \nonumber
\end{eqnarray}
Note that these expressions do not include the effect of (possibly existing) external impedance which we do not specify here. If needed, corresponding modifications can easily be implemented in a standard manner \cite{SZ}.

In addition to the above phase-phase correlation functions in what follows we will also need to define the expectation value of the current operator
\begin{equation}
\langle \hat I(t)\rangle =ie\int {\cal D} \varphi_{\pm}\frac{\delta}{\delta
  \varphi_-(t)}e^{iS_c[\varphi]+iS_t[\varphi]} \label{curr}
\end{equation}
and the current-current correlation function
$\langle \hat I(t)\hat I(t')\rangle$. For the symmetrized version of this
correlator we have
\begin{eqnarray}
&& \langle \hat I(t)\hat I(t')+\hat I(t')\hat I(t)\rangle\nonumber\\
&&=-2e^2\int {\cal D} \varphi_{\pm}\frac{\delta^2}{\delta
  \varphi_-(t)\delta\varphi_-(t')}e^{iS_c[\varphi]+iS_t[\varphi]}.
\label{corr}
\end{eqnarray}

\section{Equilibrium supercurrent noise}
We first employ our results in order to describe fluctuation effects in superconducting contacts in the absence of electron-electron interactions. In this case after performing functional derivatives with respect to $\varphi_- (t)$ in Eqs. (\ref{curr}) and (\ref{corr}) one should formally
set $\varphi_-=0$.

Let us define the noise spectrum ${\cal S}_\omega$ as
\begin{equation}
\frac12\langle \hat I(t_1)\hat I(t_2)+\hat I(t_2)\hat I(t_1)\rangle -I_S^2 =\int\limits_{-\infty}^\infty \frac{ d\omega}{2\pi}{\cal S}_\omega e^{-i \omega (t_1-t_2)}.
\end{equation}
Then from Eq. (\ref{corr}) for $\omega\ll 1/R_N C$ we easily find
\begin{equation}
{\cal S}_\omega=2e^2 {\cal I}_\omega.
\label{fnoise}
\end{equation}
Together with Eq. (\ref{iw}) this result provides the complete expression
for the equilibrium noise power spectrum in superconducting contacts with
arbitrary distribution of channel transmissions $T_n$.

In the low temperature limit $T \to 0$ and at subgap frequencies $0\leq \omega < 2\Delta$ the above general result reduces to the following expression
\begin{eqnarray}
&&{\cal S}_{\omega}=e^2\sum_n \bigg\{\frac{ \pi\Delta^2T_n^2(1-T_n)\sin^4(\chi/2)}{1-T_n\sin^2(\chi/2)}
\delta\left(\omega-2\epsilon_n(\chi)\right)
\nonumber\\
&&+T_n^{3/2}\left| \sin\frac{\chi}{2}\right|\frac{\Delta\left( \omega\epsilon_n(\chi)-\Delta^2(1+\cos\chi)\right)}{\epsilon_n(\chi)\left((\omega-\epsilon_n(\chi))^2-
\epsilon^2_n(\chi)\right)} \nonumber\\
&&\times \sqrt{(\omega-\epsilon_n(\chi))^2-\Delta^2}\theta(\omega-\Delta-\epsilon_n(\chi))
\bigg\},
\label{ztnoise}
\end{eqnarray}
where $\theta(t)$ is the Heaviside step function.  Eq. (\ref{ztnoise}) demonstrates that the contribution of each transmission channel to the noise spectrum at has a narrow peak at $\omega=2\epsilon_n(\chi)$ while at higher frequencies $\omega >\Delta+\epsilon_n(\chi)$ continuous noise spectrum sets in.
For even higher frequencies $\omega>2\Delta$ also quasiparticles with energies above the gap contribute to the noise spectrum and in the high frequency limit $\omega\gg \Delta$ Eqs. (\ref{fnoise}), (\ref{iw}) reduce to the standard Nyquist expression for normal conductors
\begin{equation}
{\cal S}_\omega \simeq \frac{\omega}{R_N} \coth\frac{\omega}{2T}.
\label{Nyquist}
\end{equation}
We also note that the expression presented in the first line of our Eq. (\ref{ztnoise}) matches with the result previously derived in Ref. \onlinecite{Madrid}.

Let us discuss some properties of the quantum low frequency current noise (\ref{ztnoise}) in more details. We observe that ${\cal S}_\omega$ essentially depends both on the channel transmission values $T_n$ and
on the phase difference $\chi$. The amplitude of the peak at the frequency $\omega=2\epsilon_n(\chi)$ increases with $T_n$ at small transmissions and decreases at higher $T_n$ vanishing in the limit
of perfect channel transmission $T_n \to 1$ except for a special point $\chi =\pi$ in which case the contribution of a fully open channel
reduces to the universal peak at zero frequency.
Combining this peak with the continuous spectrum contribution, for a fully open single channel at $T=0$ and $\chi =\pi$ we obtain
\begin{equation}
{\cal S}_{\omega}=\pi e^2\Delta^2\delta (\omega)+e^2\Delta\sqrt{1-\frac{\Delta^2}{\omega^2}}\theta(\omega-\Delta).
\end{equation}

\begin{figure}
\includegraphics[width=8cm]{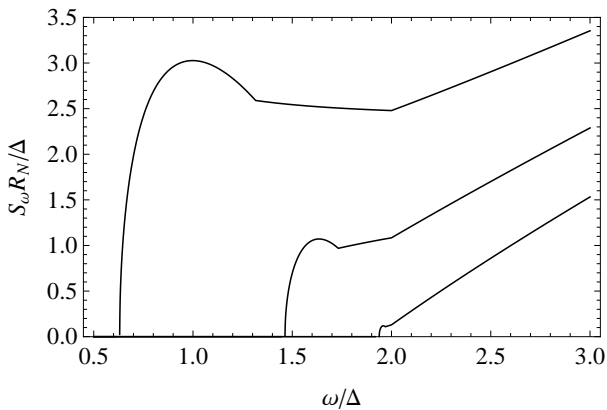}
\caption{Low temperature noise spectrum for diffusive superconducting contacts at $\chi=2.5,1.5,0.5$ (from top to bottom).}
\end{figure}

In the case of many conducting channels with different $T_n$ narrow peaks
originating from different channels occur at different frequencies and a smoother noise spectrum is observed. An important example is a diffusive conductor characterized by the the so-called bimodal transmission distribution
\begin{equation}
P(T_n)\propto \frac{1}{T_n\sqrt{1-T_n}}.
\label{bimodal}
\end{equation}
Averaging the result (\ref{ztnoise}) with this transmission distribution we arrive at the equilibrium zero temperature noise spectrum of diffusive superconducting contacts. The corresponding results are displayed in Fig. 2 for different values of the phase difference $\chi$. The noise spectrum is zero for $\omega<2\Delta |\cos (\chi/2)|$, it increases with $\omega$ at $\omega>2\Delta |\cos (\chi/2)|$ reaching the maximum at
\begin{equation}
\frac{\omega}{\Delta}=\frac{\sqrt{1+\cos\chi+\sqrt{(1+\cos\chi)(17+\cos\chi)}}}{\sqrt{2}}
\end{equation}
and showing cusps at $\omega=\Delta(1+|\cos(\chi/2)|)$ and $\omega=2\Delta$. In the limit $\omega\gg 2\Delta$ the Nyquist noise (\ref{Nyquist}) is recovered.

At non-zero temperatures there appear additional contributions to the noise spectrum. In particular, an extra peak at zero frequency emerges with the amplitude which depends on temperature as $\propto \cosh^{-2}(\epsilon_n(\chi)/2T)$, cf. Eq. (\ref{I1}). This additional thermal noise peak was previously discussed in Refs. \onlinecite{Madrid,Averin}.

Finally, we would like to point out that very recently a general analysis of persistent current noise in normal rings was
developed \cite{SZ10}. Similarly to our present findings, this analysis demonstrates that persistent current noise spectrum
has the form of sharp peaks occuring at zero frequency and at frequencies determined by the interlevel distances for
quantum states with nonzero transition matrix elements. In the low temperature limit the zero frequency peak disappears
while the peaks at non-zero $\omega$ persist down to $T=0$. Essentially the same situation is observed in superconducting contacts analyzed here.

\section{Capacitance renormalization}
 Let us now take into account small voltage fluctuations $V(t)$ across the contact. Since our present consideration is restricted to small fluctuations of the phase $\varphi_+(t)=e\int_0^t dt' V(t') \ll 1$,
 the constant in time part of the voltage should be equal to zero and the Fourier amplitude of its fluctuating part should obey the condition $eV_\omega\ll \omega$.

Under these conditions the total current $I$ across the superconducting contact takes the form
\begin{equation}
I=I_S(\chi)+\frac{C}{e}\ddot\varphi_{+}(t)-e \int d t'{\cal R}(t-t')\varphi_+(t')+\delta I(t),
\label{Langevin}
\end{equation}
where $I_S(\chi)$ is the supercurrent (\ref{Ichi}), the term involving $C$ represents the displacement current, the ${\cal R}$-dependent term accounts for the retarded current response on the fluctuating voltage $V(t)$  and $\delta I(t)$ is the stochastic contribution to the current with the correlator $\langle\delta I\delta I\rangle_\omega ={\cal S}_\omega$ studied in the previous section. Eq. (\ref{Langevin}) represents the quasiclassical Langevin equation  describing small fluctuations of the Josephson phase in superconducting contacts.

Let us analyze the Fourier amplitude $I_\omega$ of the current in the limit of small temperatures and frequencies
\begin{equation}
T, \omega\ll 2\epsilon_n(\chi ).
\label{adiabatic}
\end{equation}
We remark that the condition (\ref{adiabatic}) may yield parametrically different restrictions for weakly and highly transparent channels, cf. Eq. (\ref{And}). In the limit (\ref{adiabatic}) the noise term $\delta I$ in Eq. (\ref{Langevin}) vanishes, while the kernel ${\cal R}$ can be expanded in $\omega$ up to $\sim \omega^2$ terms. Then we obtain
\begin{eqnarray}
&& I_\omega=e\Delta\varphi_{+\omega}\sum_n \frac{T_n\left( \cos\chi + T_n\sin^4(\chi/2)\right)}{\left(1-T_n\sin^2(\chi/2) \right)^{3/2}}
\nonumber
\\ &&-\frac{C^*(\chi)}{e}\omega^2\varphi_{+\omega}. \label{lfc}
\end{eqnarray}
The first term in this expression accounts for the shift of $\chi$ in Eq. (\ref{Ichi}) by $2\varphi_{+\omega}$. The renormalized capacitance $C^*(\chi)$ involved in the second term of Eq. (\ref{lfc}) is defined as
\begin{equation}
C^*(\chi)=C+\delta C(\chi),
\label{C*}
\end{equation}
where in the limit $T\to 0$ we have
\begin{eqnarray}
&& \delta C(\chi)=\frac{e^2}{4\Delta}\sum_n \bigg\{ \frac{ 2-(2-T_n)\sin^2(\chi/2)}{T_n\sin^4(\chi/2)} \label{capren}
\\&&-\left(1-T_n\sin^2(\chi/2) \right)^{-5/2}\bigg[2T_n(T_n-2)\sin^2(\chi/2) \nonumber
\\ && +5+T_n
+\frac{2-2(1+2T_n)\sin^2(\chi/2)}{T_n\sin^4(\chi/2)}\bigg]\bigg\}.\nonumber
\end{eqnarray}

Let us analyze some important limiting cases of the above general expression for $\delta C(\chi)$. In the tunneling limit $T_n \ll 1$ this result reduces to
\begin{equation}
\delta C(\chi)=\frac{3\pi}{32 \Delta R_N}\left(1-\frac{\cos\chi}{3} \right).
\end{equation}
The first -- $\chi$-independent -- term describes the well known  capacitance renormalization in Josephson tunnel junctions due to quasiparticle tunneling \cite{SZ}. The second -- $\chi$-dependent -- term (which originates from the so-called $\beta$-terms in the AES action \cite{SZ}) is usually neglected in the literature. This approximation is justified provided Coulomb effects are pronounced, phase fluctuations are strong and, hence, $\langle \cos\chi \rangle_\chi \to 0$. Here, however, we are dealing with small phase fluctuations in which case the $\chi$-dependent terms need to be fully accounted for.

In the case of small Josephson phases $\chi\ll \pi$ Eq. (\ref{capren}) reduces to the universal expression
\begin{equation}
\delta C\simeq \frac{\pi}{16 \Delta R_N}
\label{smallchi}
\end{equation}
which remains applicable for any distribution of channel transmissions. Provided all channels are transparent, i.e. $T_n\simeq 1$, Eq. (\ref{capren}) yields
\begin{equation}
\delta C\simeq  \frac{\pi}{16 \Delta R_N \cos^4(\chi/4)} \label{ht}
\end{equation}
in the region $0<\chi<\pi$ for $\chi$ not too close to $\pi$. Starting from $\pi-\chi\sim (1-T_n)^{1/5}$  the function $\delta C(\chi )$ deviates from Eq. (\ref{ht}) and tends to
\begin{equation}
\delta C\simeq\frac{e^2}{4\Delta}\sum_n\frac{1}{(1-T_n)^{3/2}}. \label{fd}
\end{equation}
for $\chi=\pi$. In the important case of diffusive contacts averaging of Eq. (\ref{capren}) with the bimodal transmission distribution (\ref{bimodal}) yields
\begin{equation}
\delta C\simeq \frac{1.05}{\Delta R_N (\pi-\chi)^2} \label{ld}
\end{equation}
for $\pi-\chi \ll \pi$, while for small values of $\chi$ we again reproduce Eq. (\ref{smallchi}). We observe, that the renormalized capacitance (\ref{ld}) for diffusive superconducting contacts diverges as the phase difference $\chi$ approaches $\pi$. This behavior is quite natural since (i) the contribution of almost fully open channels (\ref{fd}) becomes large in this limit and (ii) many such channels are available in diffusive barriers. The behavior of the renormalized capacitance is also illustrated in Fig. 3.

\begin{figure}
\includegraphics[width=8cm]{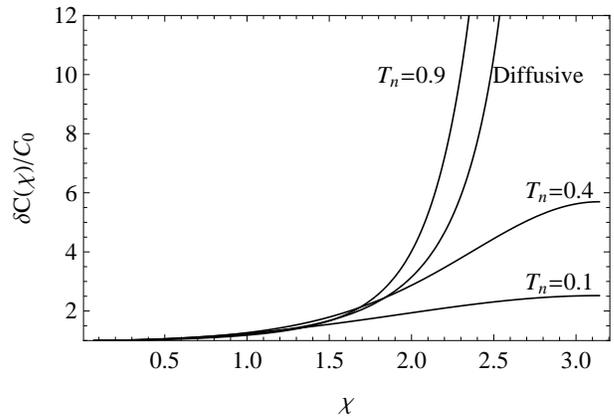}
\caption{The capacitance $\delta C(\chi)$ normalized by $C_0=\pi/16 \Delta R_N$. As indicated in the plot, three curves correspond to uniform barriers with channel transmissions $T_n=0.1,0.4, 0.9$, while the fourth curve corresponds to diffusive contacts. }
\end{figure}

Eq. (\ref{lfc}) also allows to determine the low temperature Josephson plasma frequency
$\omega_J$ of oscillations near the bottom of the Josephson potential well. We obtain
\begin{eqnarray}
\omega_J^2=\frac{e^2\Delta}{C^*(\chi)}\sum_n \frac{T_n\cos\chi + T_n^2\sin^4(\chi/2)}{(1-T_n\sin^2(\chi/2))^{3/2}}, \label{oJ}
\end{eqnarray}
where $C^*(\chi)$ is defined in Eqs. (\ref{C*}), (\ref{capren}). Strictly speaking, this expression applies only provided the condition (\ref{adiabatic}) is fulfilled. However, qualitatively it remains valid also at $\omega_J \sim 2\epsilon_n(\chi )$
up to a prefactor of order one. With this in mind, below we will employ the above expressions also in this case.

In the limit of large geometric capacitance of the junction $C \gg \delta C$ the capacitance renormalization can be neglected.
In this case we have $\omega_J\propto 1/\sqrt{C}$. In many cases, however, geometric capacitance turns out to be negligibly small
so that $C^* \simeq \delta C$. In such cases the combination $\omega_J/\Delta$ depends only on $\chi$ and on the barrier transmissions. This situation will be considered below.

For small $\chi \ll \pi$  and $T \to 0$ the Josephson current (\ref{Ichi}) reduces to  $I_S(\chi)=\pi\Delta\chi /(2e R_N)$ for
any transmission distribution. Combining this expression with Eq. (\ref{smallchi}) we get $\omega_{J}/\Delta=4$. This result
universally holds for small values of the Josephson phase. For higher values of $\chi$ the Josephson plasma frequency
becomes smaller. E.g. in the case of tunnel barriers $T_n \ll 1$ for $-\pi/2 <\chi <\pi/2$ one trivially finds
\begin{equation}
\frac{\omega_J}{\Delta}=4\sqrt{\frac{2\cos\chi}{3-\cos\chi}}.
\end{equation}

In the case of the highly transparent contacts the Josephson plasma frequency can be written as
\begin{equation}
\frac{\omega_J}{\Delta}=4\cos^2 \frac{\chi}{4}\sqrt{\cos\frac{\chi}{2}},
\end{equation}
where we assume that $-\pi<\chi<\pi$ and $\pi-|\chi |>(1-T_n)^{1/5}$.
As in this case the critical current is achieved at $\chi_{max}\simeq \pi -2(1-T_n)^{1/4}$,
for $\chi$ close to $\chi_{max}$ we obtain
\begin{equation}
\frac{\omega_J}{\Delta}=2\sqrt{2}(1-T_n)^{1/8}\sqrt{\chi_{max}-\chi}.
\end{equation}
\begin{figure}
\includegraphics[width=8cm]{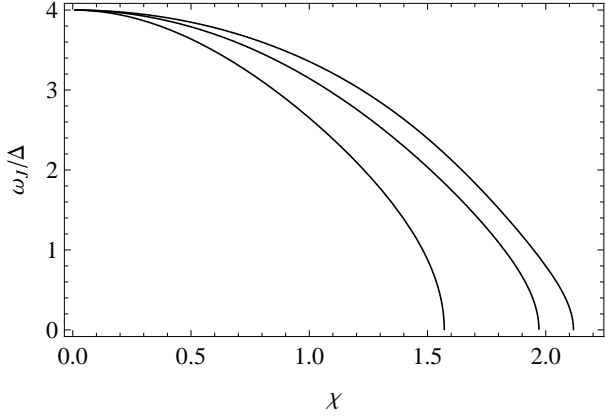}
\caption{The Josephson plasma frequencies of highly transparent ($T_n=0.9$), diffusive and tunnel barriers (from top to bottom) in the limit $\delta C \gg C$.}
\end{figure}
Provided the current is close to the critical one, the Josephson plasma frequency tends to zero as
\begin{equation}
\frac{\omega_J}{\Delta}=\gamma\left( 1-\frac{I_S^2}{I_C^2}\right)^{1/4},
\end{equation}
where $\gamma=4\sqrt{2/3}$ for tunnel barriers, $\gamma\simeq 2.53$ in the diffusive limit and $\gamma=2\sqrt{2}(1-T_n)^{1/8}$ for highly transparent junctions. The behavior of $\omega_J(\chi )$ for tunnel, diffusive and highly transparent barriers is also illustrated in Fig. 4.

Finally, let us display the adiabatic form of effective action for a
superconducting contact which remains applicable for small phase fluctuations under the condition (\ref{adiabatic}). It reads
\begin{eqnarray}
&&S_c+S_t\simeq \label{Sad}\\ &&\int\limits_0^t dt'\left[\frac{C^*(\chi)}{e^2}
\dot\varphi^+\dot\varphi^-
-\frac{I_S(\chi+2\varphi_+(t'))}{e}\varphi_-(t')\right],\nonumber
\end{eqnarray}
where $I_S(\chi )$ is defined in Eq. (\ref{Ichi}).

\section{Interaction correction to supercurrent}

Let us now turn to the electron-electron interaction correction to the equilibrium Josephson current (\ref{Ichi}). Previously such correction was analyzed in the case of Josephson tunnel barriers in the presence of linear Ohmic dissipation \cite{PZ} (see also \cite{SZ}). The task at hand is to investigate the interaction correction to the supercurrent in contacts with arbitrary transmission distribution. As before in this paper, we do not include any external impedance into our consideration.

In order to evaluate the interaction correction it is necessary to go beyond the Gaussian effective action (\ref{finalsS})-(\ref{SIII}) and to
evaluate the higher order contribution $\sim \varphi^3$. It is easy to observe that the interaction correction to the supercurrent is provided
by the following non-Gaussian terms in the effective action:
\begin{eqnarray}
&& \delta (iS_t)=
\label{3rd}\\ && \int\int\int dt_{1} dt_{2}dt_{3}Y(t_1,t_2,t_3)\varphi_-(t_1) \varphi_+(t_2)\varphi_+(t_3)\nonumber\\ && +\int\int\int dt_{1} dt_{2}dt_{3}Z(t_1,t_2,t_3)\varphi_+(t_1) \varphi_-(t_2)\varphi_-(t_3).
\nonumber
\end{eqnarray}
The function $Y(t_1,t_2,t_2)$ can be written as
\begin{eqnarray}
&& Y(t_1,t_2,t_3)=\\&&\int \int \frac{d\omega_1}{2\pi} \frac{d\omega_2}{2\pi}Y(\omega_1,\omega_2) e^{-i\omega_1(t_1-t_2)} e^{-i\omega_2(t_1-t_3)},\nonumber
\end{eqnarray}
where $Y(\omega_1,\omega_2)=Y(\omega_2,\omega_1)$. The function $Z(t_1,t_2,t_3)$ can be expressed in a similar way.

Adding the non-Gaussian terms (\ref{3rd}) to the action and employing Eq. (\ref{curr}) we arrive at the following expression for the interaction correction
\begin{eqnarray}
&& \delta I_S(\chi)=ie\int \limits_{-\infty}^{\infty} \frac{d\omega}{2\pi} Y(\omega,-\omega) \langle \varphi_+\varphi_+\rangle_\omega \nonumber\\ && +2ie\int \limits_{-\infty}^{\infty} \frac{d\omega}{2\pi} Z(0,-\omega)\langle \varphi_+\varphi_-\rangle_\omega , \label{zs}
\end{eqnarray}
where the phase-phase correlators are defined in Eq. (\ref{phiphi}).

Let us consider the first term in the right-hand side of Eq. (\ref{zs}).
It is easy to see that in the limit of low temperatures only frequencies $|\omega|>\Delta+\epsilon_n(\chi)$ contribute to the integral in Eq. (\ref{phiphi}) for $\langle \varphi_+\varphi_+\rangle$ while the
contribution from the frequency interval $|\omega|<\Delta+\epsilon_n(\chi)$ vanishes. Furthermore, the leading contribution from the first term in  Eq. (\ref{zs}) is picked up logarithmically from the interval $2\Delta\ll |\omega| \ll 1/R_N C$ where
\begin{equation}
 \langle \varphi_+\varphi_+\rangle_\omega \simeq \frac{e^2 R_N}{|\omega |}
\label{as}
\end{equation}
and the function $Y(\omega,-\omega)$ tends to a frequency independent value.

After a straightforward but tedious calculation (some relevant details are presented in Appendix) in the interesting frequency range  $\omega\gg \Delta$ from Eq. (\ref{St}) one finds
\begin{eqnarray}
&&Y(\omega,-\omega)=\frac{i\Delta\sin\chi}{4}\label{yw}\\&& \times\sum_n \frac{T_n(1-T_n)(2-T_n\sin^2(\chi/2))}{(1-T_n\sin^2(\chi/2))^{3/2}}
F(\epsilon_n(\chi)).\nonumber
\end{eqnarray}
Similarly to the third term in Eq. (\ref{currr}) this high-frequency term involves the factor $1-T_n$, i.e. it vanishes for fully open conducting channels. Combining Eqs. (\ref{as}), (\ref{yw}) with (\ref{zs}), we arrive at the expression for the supercurrent
\begin{equation}
I(\chi )=I_S(\chi )+\delta I_S(\chi).
\end{equation}
In the limit of low temperatures the interaction correction reads
\begin{eqnarray}
&& \delta I_S(\chi)=-\frac{e\Delta}{2g_N}\ln\left(\frac{1}{2\Delta R_NC} \right) \sin\chi \label{intcor}\\ && \times  \sum_n \frac{ T_n(1-T_n)}{(1-T_n\sin^2(\chi/2))^{3/2}}\left( 2-T_n\sin^2 \frac{\chi}{2}\right), \nonumber
\end{eqnarray}
where $g_N=2\pi /(e^2R_N)$ is the dimensionless normal state conductance of the contact. This result is justified as long as the Coulomb correction $\delta I_S(\chi)$ remains much smaller than the non-interacting term
$I_S(\chi )$ (\ref{Ichi}). Typically this condition requires the
dimensionless conductance to be large $g_N\gg\ln (1/2\Delta R_NC)$.

Note that Eq. (\ref{intcor}) was derived only from the first term in Eq.
(\ref{zs}). The second term in this equation involving the function
$Z(0,-\omega)$ and the correlator $\langle \varphi_+\varphi_-\rangle$ can be treated analogously. The corresponding analysis demonstrates that
the contribution of this term turns out to be smaller than that of the first term by the logarithmic factor $\sim\ln (1/2\Delta R_N C)$. Accordingly, the second term in Eq. (\ref{zs}) can be safely neglected for our purposes.

Let us emphasize again an important property of the result (\ref{intcor}): The interaction correction contains the factor $1-T_n$ and, hence, vanishes for fully open barriers. In other words, {\it no Coulomb blockade of the Josephson current is expected in fully transparent superconducting contacts.} Note that this conclusion is also consistent with numerical results in Ref. \onlinecite{Madrid2}.

The expression for the interaction correction (\ref{intcor}) can further be specified in the case of diffusive contacts. In the absence of interactions the Josephson current in such contacts follows from (\ref{Ichi}) and takes the well known form
\begin{equation}
I_S(\chi)=\frac{\pi \Delta}{2 e R_N}\cos\frac{\chi}{2}\ln\frac{1+\sin\frac{\chi}{2}}
{1-\sin\frac{\chi}{2}}.
\end{equation}
Including interactions and averaging (\ref{intcor})
with the bimodal transmission distribution (\ref{bimodal}) one finds
\begin{eqnarray}
&& \delta I_S(\chi)=-\frac{e}{8}\Delta \ln\left( \frac{1}{2\Delta R_N C}\right)
\cot (\chi/2) \label{intdif}
\\&&\times \left[\left(\sin\frac{\chi}{2}+\sin^{-1}\frac{\chi}{2}\right)
\ln\frac{1+\sin(\chi/2)}{1-\sin(\chi/2)}-2\right].
\nonumber
\end{eqnarray}

Note that the result (\ref{intcor}) can formally be reproduced if one substitutes $T_n\to T_n+\delta T_n$ into
Eq. (\ref{Ichi}), where
\begin{equation}
\delta T_n=-\frac{2}{g_N}\ln \left(\frac{1}{2\Delta R_NC} \right) T_n(1-T_n),
\label{tcor}
\end{equation}
and then expands the result to the first order in $\delta T_n$.
Interestingly, the same transmission renormalization (\ref{tcor}) follows from the renormalization group (RG) equations \cite{KN,GZ04}
\begin{equation}
\frac{ d T_n}{d L}=-\frac{ T_n(1-T_n)}{\sum_k T_k},\quad L=\ln \left(\frac{1}{\epsilon R_N C} \right)
\end{equation}
derived for {\it normal} conductors. In order to arrive at Eq. (\ref{tcor}) one should just start the RG flow at $\epsilon=1/R_NC$ and stop it at $\epsilon=2\Delta$. Thus, the result (\ref{intcor}) can be interpreted in a very simple manner: Coulomb interaction provides high frequency renormalization $T_n+\delta T_n$ (\ref{tcor}) of the barrier transmissions which should be substituted into the classical expression for the supercurrent (\ref{Ichi}).
It should be stressed, however, that the last step would by no means appear obvious without our rigorous derivation since the Coulomb correction to the Josephson current originates from the term $\sim \varphi_-\varphi_+^2$ in the effective action which is, of course, totally absent in the normal case.

\section{Discussion}
The analysis employed in this paper demonstrates that fluctuation and
interaction effects in superconducting contacts with arbitrary
transmissions of conducting modes show a number of qualitatively new features
as compared to the case of Josephson tunnel barriers \cite{SZ}.
The main physical reason behind such differences is the presence of subgap
Andreev bound states (\ref{And}) inside the system.

In the limit of sufficiently small fluctuations of the Josephson phase
difference we derived the complete expression for the effective action of
superconducting contacts. This expression allowed us to obtain the general
result for the equilibrium current-current correlation function describing
Josephson current noise in such contacts. Due to the presence of subgap bound
states this current noise essentially depends on the Josephson phase $\chi$
and remains non-zero even in the zero temperature limit and at subgap
frequencies. For instance, in a physically important case of diffusive
contacts at $T=0$ the equilibrium current noise spectrum differs from zero
at all frequencies exceeding the threshold value $2\Delta|\cos(\chi/2)|$
and has the form displayed in Fig. 2.

Another important effect studied here is capacitance renormalization. While
geometric capacitance of superconducting contacts can be small, retardation
effects yield additional ``capacitance-like''
contributions which depend on both the transmission distribution $T_n$ and the
Josephson phase $\chi$ and can well exceed the geometric capacitance term. In
particular, in the limit $\chi \to \pi$ the renormalized capacitance was
found to diverge in highly transparent and diffusive contacts, see Fig. 3.
This behavior differs from that of tunnel junctions \cite{SZ} and
can also substantially affect the frequency of Josephson plasma oscillations,
see Fig. 4.

Finally, our effective action formalism enabled us to analyze the correction
to the Josephson current due to electron-electron interactions. Provided this
Coulomb correction $\delta I_C$ remains small as compared
to the Josephson critical current $I_C$ one finds the universal result
\begin{equation}
\frac{\delta I_C}{I_C}=-\frac{\alpha}{g_N}\ln\left(\frac{1}{2\Delta R_N C}\right),\label{intst}
\end{equation}
where the numerical prefactor $\alpha$ depends on the transmission
distribution. This prefactor reaches its maximum value
$\alpha=2$ in the case of tunnel barriers and becomes smaller for higher
transmissions, e.g. $\alpha\simeq 0.72$ for diffusive contacts. In the case
of fully open barriers the prefactor $\alpha$ tends to zero, $\alpha \to 0$,
implying that no Coulomb blockade of the Josephson current is expected in such barriers.

Our results for the interaction correction, e.g., Eq.
(\ref{intst}), might explain a rapid change between superconducting and
insulating behavior recently observed  \cite{Bezr} in
comparatively short metallic wires with resistances close to
the quantum resistance unit $\sim 6.5$ K$\Omega$ in-between two bulk
superconductors. Previously it was already
argued \cite{AGZ} that such a superconductor-to-insulator crossover
can be due to Coulomb effects. Our present results provide further
quantitative arguments in favor of this conclusion.

Let us again stress that in this paper we only
considered the limit of relatively short contacts in which case the
characteristic Thouless energy  of the contact $\epsilon_{\rm Th}$ exceeds the
superconducting order parameter $\Delta$. In the opposite case of long
junctions $\epsilon_{\rm Th}\ll \Delta$ (which appears to be more relevant to the
experiments \cite{Bezr}) the interaction correction to the supercurrent turns
out to have the same structure (\ref{intst})
with $2\Delta$ substituted by $\epsilon_{\rm Th}$ (see also \cite{GZ06}). The
corresponding analysis will be published elsewhere.

\vspace{0.5cm}

\centerline{\bf Acknowledgment}

\vspace{0.5cm}

This work was supported in part by RFBR grant 09-02-00886.

\appendix

\section{}
Let us present our results for the kernels ${\cal I}(t)$ and ${\cal R}(t)$ in Eqs. (\ref{SRRR}), (\ref{SIII}). The Fourier transform ${\cal I}_\omega$ of the kernel ${\cal I}(t)$ is even in $\omega$, real and non-negative function. Hence, ${\cal I}(t)$ is a real function. For $\omega\ge 0$ we have
\begin{widetext}
\begin{eqnarray}
&& {\cal I}_\omega=\sum_n \left\{  \frac{\pi T_n^2}{8\epsilon_n^2(\chi)}\Delta^4\sin^2\chi\left[1-F^2(\epsilon_n(\chi))\right]\delta(\omega)+
\frac{\pi T_n^2(1-T_n)}{4\epsilon^2_n(\chi)}\Delta^4\sin^4\frac{\chi}{2}\left[1+F^2(\epsilon_n(\chi))\right]
 \delta\left(\omega-2\epsilon_n(\chi)\right) \right. \label{iw}
\\
 && +T_n^{3/2}\left[1-F(\epsilon_n(\chi)) F(\omega+\epsilon_n(\chi)) \right]\theta(\omega-\Delta+\epsilon_n(\chi))\left| \sin\frac{\chi}{2}\right|\frac{\Delta\left( \omega\epsilon_n(\chi)+\Delta^2(1+\cos\chi)\right)}{4\epsilon_n(\chi)\left((\omega+\epsilon_n(\chi))^2-
\epsilon^2_n(\chi)\right)}\sqrt{(\omega+\epsilon_n(\chi))^2-\Delta^2} \nonumber\\
&& +T_n^{3/2}\left[1+F(\epsilon_n(\chi)) F(\omega-\epsilon_n(\chi)) \right]\theta(\omega-\Delta-\epsilon_n(\chi))\left| \sin\frac{\chi}{2}\right|\frac{\Delta\left( \omega\epsilon_n(\chi)-\Delta^2(1+\cos\chi)\right)}{4\epsilon_n(\chi)\left((\omega-\epsilon_n(\chi))^2-
\epsilon^2_n(\chi)\right)}\sqrt{(\omega-\epsilon_n(\chi))^2-\Delta^2} \nonumber
\\&& +T_n\int\limits_\Delta^\infty \frac{d\epsilon}{2\pi}\frac{\sqrt{\epsilon^2-\Delta^2}\sqrt{(\epsilon+\omega)^2-\Delta^2}}{
(\epsilon^2-\epsilon^2_n(\chi))((\epsilon+\omega)^2-\epsilon^2_n(\chi))}\left( \epsilon(\omega+\epsilon)+\Delta^2\cos\chi+T_n\Delta^2\sin^2\frac{\chi}{2}\right)\left( 1-F(\epsilon)F(\epsilon+\omega)\right)+\nonumber
\\ && \left. \frac{T_n}{2}\theta(\omega-2\Delta)\int\limits_\Delta^{\omega-\Delta} \frac{d\epsilon}{2\pi} \frac{\sqrt{\epsilon^2-\Delta^2}\sqrt{(\omega-\epsilon)^2-\Delta^2}}{
(\epsilon^2-\epsilon^2_n(\chi))((\omega-\epsilon)^2-\epsilon_n^2(\chi))}\left( \epsilon(\omega-\epsilon)-\Delta^2\cos\chi-T_n\Delta^2\sin^2\frac{\chi}{2}\right)\left( 1+ F(\epsilon)F(\omega-\epsilon)\right)\right\}.\nonumber
\end{eqnarray}
\end{widetext}

The Fourier transform of the function ${\cal R}(t)$ can be written in the form
${\cal R}_\omega={\cal R}'_\omega+i {\cal R}''_\omega$, where the real functions ${\cal R}'_\omega$ and ${\cal R}''_\omega$ are respectively even and odd in $\omega$. Hence, ${\cal R}(t)$ is also a real function. The function ${\cal R}''_\omega$  is linked to ${\cal I}$ by means of FDT relation(\ref{FDT}), and the function ${\cal R}'_\omega$ for $\omega>0$ reads
\begin{widetext}
\begin{eqnarray}
&& {\cal R}'_\omega=\sum_n \left\{ T_n(1-T_n)\frac{F(\epsilon_n(\chi))}{\epsilon_n(\chi)}\Delta^2\sin^2\frac{\chi}{2}-2\frac{T_n^2(1-T_n) F(\epsilon_n(\chi))}{(\omega^2-4\epsilon^2_n(\chi))\epsilon_n(\chi)}\Delta^4\sin^4\frac{\chi}{2} \right. \label{im1}\\
&&+2T_n\int\limits_{\Delta-{\rm min}\{\omega,2\Delta\}}^{\Delta}\frac{d\epsilon}{2\pi}
\frac{F(\epsilon+\omega)\left( \epsilon(\omega+\epsilon)+\Delta^2\cos\chi +T_n\Delta^2\sin^2(\chi/2)\right)}{(\epsilon^2-\epsilon^2_n(\chi))((\epsilon+\omega)^2-\epsilon^2_n(\chi))}
\sqrt{\Delta^2-\epsilon^2}\sqrt{(\epsilon+\omega)^2-\Delta^2}
\nonumber
\\ && +\frac{T_n^{3/2}\Delta|\sin(\chi/2)|F(\epsilon_n(\chi))}{2\epsilon_n(\chi)}\left[ \theta(\Delta-\epsilon_n(\chi)-\omega) \frac{\Delta^2(1+\cos\chi)+\omega\epsilon_n(\chi)}{(\omega+\epsilon_n(\chi))^2-\epsilon^2_n(\chi)}
\sqrt{\Delta^2-(\omega+\epsilon_n(\chi))^2}+\right.\nonumber
\\ && \left.\left. +\theta(\Delta+\epsilon_n(\chi)-\omega) \frac{\Delta^2(1+\cos\chi)-\omega\epsilon_n(\chi)}{(\omega-\epsilon_n(\chi))^2-\epsilon^2_n(\chi)}
\sqrt{\Delta^2-(\omega-\epsilon_n(\chi))^2}\right]\right\}. \nonumber
\end{eqnarray}
\end{widetext}
The integral in this expression is taken in the v.p. sense. The functions ${\cal R}'_\omega$ and ${\cal R}''_\omega$ are also related by the causality principle: the function ${\cal R}(t)$ should vanish for $t<0$.
E.g., the contribution of Andreev bound states to ${\cal R}''_\omega$ has
the form
\begin{eqnarray}
&& \pi T_n^2(1-T_n) \frac{F(\epsilon_n(\chi))}{\epsilon^2_n(\chi)}\Delta^4\sin^4 \frac{\chi}{2} \left[ \delta(\omega-2\epsilon_n(\chi))\right.\nonumber
\\ && \left.-\delta(\omega+2\epsilon_n(\chi))\right].
\end{eqnarray}
It contributes to ${\cal R}(t)$ as
\begin{equation}
T_n^2(1-T_n) \frac{F(\epsilon_n(\chi))}{\epsilon^2_n(\chi)} \Delta^4\sin^4\frac{\chi}{2}\sin[2\epsilon_n(\chi) t]. \label{fc}
\end{equation}
The corresponding contribution from discrete Andreev states to
${\cal R}'_\omega$ reads
\begin{equation}
-4\frac{T_n^2(1-T_n) F(\epsilon_n(\chi))}{(\omega^2-4\epsilon^2_n(\chi))
\epsilon_n(\chi)}\Delta^4\sin^4\frac{\chi}{2}. \label{sqd}
 \end{equation}
One half of this contribution comes from the second term in Eq. (\ref{im1}), the other half stems from the second Heaviside function in Eq. (\ref{im1}). Making use of the integral
\begin{equation}
\int\limits_0^\infty dx \frac{\cos (b x)}{x^2-y^2}=-\frac{\pi}{2y}\sin(|b|y),
\end{equation}
taken in the v.p. sense, we observe, that the contribution ({\ref{fc}}) increases by the factor two for $t>0$ and it vanishes for $t<0$. Hence, the total contribution of Andreev levels to ${\cal R}(t)$ is given by the last term of Eq. (\ref{currr}).

One can also write
\begin{eqnarray}
&& {\cal R}(t)=\sum_n\left\{ -\frac{2}{\pi} \theta(t)\int_0^\infty d\omega \sin(\omega t)\left(  f_\omega+\frac{T_n}{\pi}\omega \right)\right. \nonumber
\\ && -\frac{T_n}{\pi}\delta'(t)+T_n(1-T_n)\frac{F(\epsilon_n(\chi))}{\epsilon_n(\chi)} \Delta^2\sin^2\frac{\chi}{2}\delta(t) \label{currr}
\\ && \left. +2T_n^2(1-T_n) \frac{F(\epsilon_n(\chi))}{\epsilon^2_n(\chi)} \Delta^4\sin^4\frac{\chi}{2}\theta(t)\sin(2\epsilon_n(\chi) t)\right\},\nonumber
\end{eqnarray}
where
\begin{widetext}
\begin{eqnarray}
&& f_\omega= 2 T_n\int\limits_\Delta^\infty \frac{d\epsilon}{2\pi}\frac{\sqrt{\epsilon^2-\Delta^2}\sqrt{(\epsilon+\omega)^2-\Delta^2}}{
(\epsilon^2-\epsilon^2_n(\chi))((\epsilon+\omega)^2-\epsilon^2_n(\chi))}\left( \epsilon(\omega+\epsilon)+\Delta^2\cos\chi+T_n\Delta^2\sin^2\frac{\chi}{2}\right)\left( F(\epsilon)-F(\epsilon+\omega)\right)\label{fw}
\\ && -T_n\theta(\omega>2\Delta)\int\limits_\Delta^{\omega-\Delta} \frac{d\epsilon}{2\pi} \frac{\sqrt{\epsilon^2-\Delta^2}\sqrt{(\omega-\epsilon)^2-\Delta^2}}{
(\epsilon^2-\epsilon^2_n(\chi))((\omega-\epsilon)^2-\epsilon^2_n(\chi))}\left( \epsilon(\omega-\epsilon)-\Delta^2\cos\chi-T_n\Delta^2\sin^2\frac{\chi}{2}\right)\left( F(\epsilon)+F(\omega-\epsilon)\right)\nonumber
\\ && -\frac{ T_n^{3/2}\Delta}{2\epsilon_n(\chi)}\left| \sin\frac{\chi}{2}\right|\left[ \frac{ \omega\epsilon_n(\chi)+\Delta^2(1+\cos\chi)}{(\omega+\epsilon_n(\chi))^2-
\epsilon^2_n(\chi)}\sqrt{(\epsilon_n(\chi)+\omega)^2-\Delta^2}\left[  F(\omega+\epsilon_n(\chi))- F(\epsilon_n(\chi)) \right]\theta(\omega-\Delta+\epsilon_n(\chi)) \right. \nonumber
\\ &&\left.+ \frac{ \omega\epsilon_n(\chi)-\Delta^2(1+\cos\chi)}{(\omega-\epsilon_n(\chi))^2-
\epsilon^2_n(\chi)}\sqrt{(\omega-\epsilon_n(\chi))^2-\Delta^2}\left[  F(\omega-\epsilon_n(\chi))+ F(\epsilon_n(\chi)) \right]\theta(\omega-\Delta-\epsilon_n(\chi)) \right].\nonumber
\end{eqnarray}
\end{widetext}
The second term in Eq. (\ref{currr}) describes Ohmic damping and the third contribution represents the supercurrent correction. Note that this correction turns out to be proportional to the factor $T_n(1-T_n)$. Finally, the last term accounts for the effect of discrete Andreev levels as it is described above.

Setting $\epsilon_{m}={\rm max}\{\Delta,T\}$, for $\epsilon_{m} t\ll 1$ with the logarithmic accuracy we obtain
\begin{eqnarray}
&& \frac{2}{\pi} \theta(t)\int\limits_0^\infty d\omega \sin(\omega t)\left(  f_\omega+\frac{T_n}{\pi}\omega \right)\simeq\\ &&\frac{2}{\pi} \theta(t) T_n\left( \Delta^2\cos\chi + T_n \Delta^2\sin^2\frac{\chi}{2}\right)\ln\frac{1}{t \epsilon_{m}}.\nonumber
\end{eqnarray}

Let us also present some details relevant for the calculation of the kernel $Y(\omega,-\omega)$ in the limit $\omega\gg \Delta$. This kernel can be represented as a sum
\begin{equation}
Y=Y^{(1)}+Y^{(2)}+Y^{(3)},\label{y123}
\end{equation}
where $Y^{(1,2,3)}$ define respectively the first, second and third order terms in the expansion of the logarithm in Eq. (\ref{xx}). They read
\begin{eqnarray}
&& Y^{(1)}=\frac{iT_n}{2}\frac{F(\epsilon_n(\chi))}{\epsilon_n(\chi)}\Delta^2\sin\chi,
\\&& Y^{(2)}=-\frac{iT_n^2}{2}\frac{F(\epsilon_n(\chi))}{\epsilon_n(\chi)}\Delta^2\sin\chi \nonumber
\\&&+\frac{iT_n^2}{8}\frac{F(\epsilon_n(\chi))}{\epsilon^3_n(\chi)}\Delta^2
\sin\chi(\epsilon^2_n(\chi)-\Delta^2\cos\chi),\nonumber
\\&& Y^{(3)}=-\frac{iT_n^3}{8}\frac{F(\epsilon_n(\chi))}{\epsilon^3_n(\chi)}\Delta^4\sin\chi \sin^2\frac{\chi}{2}.\nonumber
\end{eqnarray}
Combining these expressions with Eq. (\ref{y123})  we arrive at Eq. (\ref{yw}). Consider, e.g., a typical summand
\begin{widetext}
\begin{eqnarray}
&& \delta(i S_t)=\frac{i T_n^3}{256}\int \int \int dt_1 dt_2 dt_3 \varphi_-(t_1) \varphi_+(t_2)\varphi_+(t_3) {\rm Tr}\left[\left(\check O\check X_0^{-1}\check Q_R\right)_{t_1,t_2}\left( \check\tau_3 \check Q_L\check X_0^{-1} \check Q_R \right)_{t_2,t_3}  \left( \check\tau_3 \check Q_L\check X_0^{-1} \check Q_R \check Q_L\right)_{t_3,t_1} \right.\nonumber
\\ && \left.-\left(\check O\check Q_L \check Q_R\check X_0^{-1}\check Q_R\right)_{t_1,t_2} \left( \check \tau_3 \check Q_L \check X_0^{-1}\check Q_R\right)_{t_2,t_3}\left( \check \tau_3\check Q_L\check X_0^{-1}\right)_{t_3,t_1}\right]\label{lconv}
\end{eqnarray}
which emerges in the third order. Here the matrices $\check Q_{L,R}$  are taken for $\varphi_\pm=0$ but for non-zero $\chi$, the indices stand for the temporal arguments of the convolutions and
\begin{equation}
\check \tau_3=\left( \begin{array}{cc} \hat \tau_3 & 0\\ 0 &\hat\tau_3 \end{array}\right),\quad \check O=\left( \begin{array}{cc}0&  \hat \tau_3 \\ \hat\tau_3 & 0\end{array}\right).
\end{equation}
The contribution of Eq. (\ref{lconv}) to $Y^{(3)}(\omega,-\omega)$ reads
\begin{equation}
\delta Y^{(3)}(\omega,-\omega)=\frac{i T_n^3}{256} \int\frac{d\epsilon}{2\pi} {\rm Tr}\left[ \left( \check Q_L \check X_0^{-1}\check Q_R\check Q_L\check O\check X_0^{-1}\check Q_R -\check Q_L \check X_0^{-1}\check O\check Q_L\check Q_R\check X_0^{-1}\check Q_R\right)_{\epsilon+\omega} \left( \check\tau_3 \check Q_L \check X_0^{-1} \check Q_R \check \tau_3\right)_\epsilon\right].\label{ytyp}
\end{equation}
\end{widetext}
Taking the limit $\omega\gg \Delta$ and using the property
\begin{eqnarray}
 &&\delta (x-b)\left( \frac{1}{(x-i\delta)^2-b^2} +\frac{1}{(x+i\delta)^2-b^2}\right)\nonumber
 \\ && =-\frac{1}{2b^2}\delta(x-b),
\end{eqnarray}
we find
\begin{equation}
\delta Y^{(3)}=\frac{iT_n^3 F(\epsilon_n(\chi)}{128\epsilon_n^3(\chi)}\Delta^2\sin\chi\left(\Delta^2\cos\chi-\epsilon^2_n(\chi)\right).
\end{equation}
Depending on the way of counting there are about 50 contributions to $iS_t$ similar to Eq. (\ref{lconv}). The calculation of the corresponding contributions to $Y$ is analogous to that presented above.


\begin{thebibliography}{99}
\bibitem{BJ} B. Josephson, Phys. Lett. A {\bf 1}, 251 (1962).
\bibitem{AB} V. Ambegaokar and A. Baratoff, Phys. Rev. Lett. {\bf
10}, 486 (1963); {\bf 11}, 104 (1963).
\bibitem{KO} I.O. Kulik and A.N. Omel'yanchuk, Fiz. Nizk. Temp. {\bf
4}, 296 (1978) [Sov. J. Low Temp. Phys. {\bf 4}, 142 (1978)]; W. Haberkorn, H. Knauer, and J. Richter, Phys. Stat. Solidi (A) {\bf 47}, K161 (1978);
C.W.J. Beenakker, Phys. Rev. Lett. {\bf 67}, 3836 (1991).
\bibitem{SNScl} I.O. Kulik, Zh. Eksp. Teor. Fiz. {\bf 57}, 1745 (1969)
[Sov. Phys. JETP {\bf 30}, 944 (1970)]; C. Ishii, Progr. Theor. Phys. {\bf 44}, 1525 (1970); A.V. Galaktionov and A.D. Zaikin, Phys. Rev. B {\bf 65}, 184507 (2002).
\bibitem{SNSd} A.D. Zaikin and G.F. Zharkov, Fiz. Nizk. Temp. {\bf 7}, 375 (1981) [Sov. J. Low Temp. Phys. {\bf 7}, 184 (1981)]; F.K. Wilhelm,
   A.D. Zaikin and G. Sch\"on, J. Low Temp. Phys. {\bf 106}, 305 (1997);
   P. Dubos, H. Courtois, B. Pannetier, F.K. Wilhelm,
   A.D. Zaikin and G. Sch\"on, Phys. Rev. B {\bf 63},  064502 (2001).
\bibitem{AR} A.F. Andreev, Zh. Eksp. Teor. Fiz. {\bf 46}, 1823 (1964) [Sov. Phys. JETP {\bf 19}, 1228 (1964)].
\bibitem{lam} C.J. Lambert and R. Raimondi, J. Phys.
Cond. Mat. {\bf 10}, 901 (1998).
\bibitem{bel} W. Belzig, F.K. Wilhelm, C. Bruder, G. Sch\"on, and A.D. Zaikin,
 Superlatt. Microstr. {\bf 25}, 1251 (1999).
 \bibitem{SaMiZhe} A.A. Golubov, M.Yu. Kupriyanov, and E. Il'ichev,
  Rev. Mod. Phys. {\bf 76}, 411 (2004).
\bibitem{Madrid} A. Martin-Rodero, A. Levy Yeyati, and F.J.
Garcia-Vidal, Phys. Rev. B {\bf 53}, R8891 (1996).
\bibitem{Averin} D. Averin and H.T. Imam, Phys. Rev. Lett. {\bf
76}, 3814 (1996).
\bibitem{Yip} S. Yip,  Phys. Rev. B {\bf 68}, 024511 (2003).
\bibitem{SZ} G. Sch\"on and A.D. Zaikin, Phys. Rep. {\bf 198}, 237 (1990).
\bibitem{GZ01} D.S. Golubev and A.D. Zaikin, Phys.
Rev. Lett. {\bf 86}, 4887 (2001).
\bibitem{GGZ03} A.V. Galaktionov, D.S. Golubev, and A.D. Zaikin, Phys. Rev. B {\bf 68}, 085317 (2003); {\it ibid.} {\bf 68}, 235333 (2003).
\bibitem{KN} M. Kindermann and Yu.V. Nazarov, Phys. Rev. Lett. {\bf 91}, 136802 (2003); D.A. Bagrets and Yu.V. Nazarov, Phys. Rev. Lett. {\bf 94}, 056801 (2005).
\bibitem{GZ04} D.S. Golubev and A.D. Zaikin, Phys. Rev. B {\bf 69}, 075318 (2004); D.S. Golubev, A.V. Galaktionov, and A.D. Zaikin, Phys. Rev. B {\bf 72}, 205417 (2005).
\bibitem{GZ06} A.V. Galaktionov and A.D. Zaikin, Phys. Rev. B {\bf 73}, 184522 (2006).
\bibitem{GZ09} A.V. Galaktionov and A.D. Zaikin, Phys. Rev. B {\bf 80},
174527 (2009).
\bibitem{Z} A.D. Zaikin, Physica B \textbf{203}, 255 (1994).
\bibitem{SN} I. Snyman and Yu.V. Nazarov, Phys. Rev. B
{\bf 77}, 165118 (2008).
\bibitem{SZ10} A.G. Semenov and A.D. Zaikin, arXiv:1002.3104.
\bibitem{PZ} A.D. Zaikin and S.V. Panyukov, JETP Lett. {\bf 43}, 670 (1986); S.V. Panyukov and A.D. Zaikin, Physica B \textbf{152}, 162 (1988).
\bibitem{Madrid2} A. Levy Yeyati, J.C. Cuevas, and A. Martin-Rodero, Phys. Rev. Lett. {\bf 95}, 056804 (2005).
\bibitem{Bezr} A.T. Bollinger, A. Rogachev, and A. Bezryadin, Europhys. Lett., {\bf 76}, 505 (2006); A.T. Bollinger, R.C. Dinsmore III, A. Rogachev, and A. Bezryadin, Phys. Rev. Lett. {\bf 101}, 227003 (2008).
\bibitem{AGZ} K.Yu. Arutyunov, D.S. Golubev, and A.D. Zaikin,
Phys. Rep. {\bf 464}, 1 (2008).

\end{thebibliography}
\end{document}